\documentclass[review]{elsarticle}

\usepackage{color}
\usepackage{epsfig}
\usepackage{amsmath, amsthm}
\usepackage{mathtools}  
\mathtoolsset{showonlyrefs}
\usepackage{amssymb}
\usepackage{amscd}
\usepackage{threeparttable}
\usepackage{graphicx}
\usepackage{amssymb}
\usepackage{amscd}
\usepackage{threeparttable}
\usepackage{graphicx}

\providecommand{\norm}[1]{\lVert#1\rVert}

\journal{Systems \& Control Letters}

\begin{document}

\begin{frontmatter}

\title{Adaptive Synchronization of Heterogeneous Multi-Agent Systems: A Free Observer Approach}

\author[mymainaddress]{Miguel F. Arevalo-Castiblanco*}\cortext[mycorrespondingauthor]{Corresponding author}
\ead{miarevaloc@unal.edu.co}

\author[mymainaddress]{Duvan Tellez-Castro}\cortext[acknowledgement]{Duvan Tellez-Castro was supported by Colombian  Administrative National Department of Science, Technology, and Innovation (COLCIENCIAS) under Grant No. 727/2016.}
\ead{datellezc@unal.edu.co}

\author[mysecondaryaddress]{Jorge Sofrony}
\ead{jsofronye@unal.edu.co}

\author[mymainaddress]{Eduardo Mojica-Nava}
\ead{eamojican@unal.edu.co}

\address[mymainaddress]{Department of Electrical and Electronic Engineering, Universidad Nacional de Colombia}
\address[mysecondaryaddress]{Department of Mechanical and Mechatronic Engineering, Universidad Nacional de Colombia}

\begin{abstract}
Adaptive synchronization protocols for heterogeneous multi-agent network are investigated. The interaction between each of the agents is carried out through a directed graph. We highlight the lack of communication between agents and the presence of uncertainties in each system among the conventional problems that can arise in cooperative networks. Two methodologies are presented to deal with the uncertainties: A strategy based on robust optimal control and a strategy based on neural networks. Likewise, an input estimation methodology is designed to face the disconnection that any agent may present on the network. These control laws can guarantee synchronization between agents even when there are disturbances or no communication from any agent. Stability and boundary analyzes are performed. Cooperative cruise control simulation results are shown to validate the performance of the proposed control methods.
\end{abstract}

\begin{keyword}
Adaptive control \sep Distributed control \sep Optimal modification \sep Neural network approximation \sep Uncertainty management
\MSC[2020] 00-01\sep  99-00
\end{keyword}

\end{frontmatter}

\section{Introduction} \label{S1}

Nowadays, multi-agent system control has a wide variety of applications ranging from mechanical systems like robots synchronization to social behavior for influence analysis \cite{Cao2012,Luis2020,Nguyen2020}. For distributed control systems, problems based on a reference or leader model have been widely considered \cite{Olfati2006,Zhu2019, Zhao2019, Zhao20192}. Cooperative cruise control has drawn great attention as an application case among the main problems to be tackled \cite{Dey2015,Gong2019}.

Conventional control techniques for cooperative cruise systems have been used for more than 20 years with effective results \cite{Stanton1997,Prestl2000}. However, the systems are considered without disturbances neither network communication failures, limiting the application of these control methods in practice.

Adaptive control theory is proposed as an effective method for dynamic systems with uncertainty parameters \cite{Hou2009,Nguyen2008}, such as Model Reference Adaptive Control (MRAC) for leader-follower models. MRAC allows the online adjustment of the controller parameters through adaptive laws in order to synchronize a dynamic with respect to a reference model \cite{Parks1966}. For distributed systems, MRAC is extended including matching conditions for both, the reference and the dynamics of neighboring agents \cite{Baldito2018}. Similarly, as a robust complement, optimal adaptive theory or neural network approach could be included to mitigate input system uncertainties \cite{Arevalo2019,Arevalo20192}.

Likewise on a cooperative practical level, one of the common scenarios to be presented is the communication loss between agents \cite{Li2009}. In the distributed MRAC case, where communication of each agent control input is handled, the lack of communication of this variable is common in practice due to different conditions such as disconnection from the network by physical environment or by pre-established configuration against energy losses. The conventional control strategies present failures in their operation by not considering a lack of communication \cite{Qin2013}, or handle it as a disturbance but without considering an adaptive law for estimation \cite{Peng2013}. The challenge is build some adaptive protocols to allow an input estimation of uncommunicated agents and also addresses input uncertainties in the case of a heterogeneous agents network.

In recent works, just few results have been proposed using the MRAC from a robust optimal perspective to handle disturbances in centralized way \cite{Nguyen2008,Arabi2019}, and some others results have proposed on distributed input estimation, but without covering additional uncertainty parameters \cite{Baldi2019, Shen2019}. In the distributed control used to synchronize heterogeneous agents, the controller must adjust four sets of parameters: the feedback matching conditions related to the dynamics of the reference agent, the coupling matching conditions related to the dynamics of neighboring agents, the uncertainty optimal parameters for the suppression of disturbances, and the input estimation parameter for the uncommunicated agents input. These parameters should be adjusted for agents that are directly communicated with the reference and those which are not \cite{Baldito2018,Arevalo2019}.

The main contribution of this paper is Threefold, first the development of a control protocol that allows the suppression of uncertainty parameters, second, the development of a law for estimating the input in cases of communication failure, and third the conjunction of these theories into a general control protocol. The validation of established control laws is presented through a boundary analysis using Lyapunov's theory. In particular, a third order cooperative cruise control simulation case is presented to show the improvement in the temporal response of the implemented control laws.

The rest of the paper is organized as follows. Section II presents the formulation of the problem and the mathematical preliminaries, in Section III the development of control laws for managing uncertainties through adaptive optimal theory and the use of neural networks are presented. Section IV presents the development of the control law for estimating input parameters, in Section V the simulation results under the cooperative cruise control study case is presented, and finally section VI presents the conclusions of the work done and the projections of future work.

\section{Problem Formulation and Mathematical Preliminaries} \label{S2}

This section shows a contextualization of the heterogeneous agents  synchronization problem and the basic notations to be used throughout the paper.

Consider a heterogeneous network with $N$ agents and a reference model. Communication presented in the network is defined by a graph $\mathcal{G}=(\mathcal{V},\mathcal{E})$, where $\mathcal{V}=\{1,2,\ldots,N\}$ is the set of nodes or agents in the network and $\mathcal{E}\subseteq\mathcal{V}\times\mathcal{V}$ is the link set of $\mathcal{G}$. To determine communication, if there is a link between agent $i$ and agent $j$ then $(i,j)\in\mathcal{E}$, which means that $i$ and $j$ are neighboring agents. For agent synchronization, the synchronization error between agent $i$ and agent $j$ is defined as $e_{ij}=x_i-x_j$. Let $\mathcal{A}=[a_{ij}]\in\mathbb{R}^N$ be the adjacency matrix of the graph where $a_{ij}>0$ if $(i,j)\in\mathcal{E}$, $a_{ij}=0$ if $(i,j)\not\in\mathcal{E}$, and $a_{ii}=0\;\forall i$. When a direct succession of paths is found from an agent $i$ to an agent $j$, $\{(i,k),(k,l),\ldots,(l,j)\}$ is defined as a direct path. If there is a direct path for each agent it is defined as spanning tree. The Laplacian of a graph is denoted as $L=D-\mathcal{A}$ where $D=\text{diag}\{d_1,d_2,\ldots,d_N\}$, with $d_i=\sum_{j=1}^Na_{ij}$. 
Matrices and vectors it is used $X$ and $x$ respectively. The transpose of a matrix or vector are defined as $X^\top$ and $x^\top$ respectively. The inverse of a square matrix is represented as $A^{-1}$ and the transpose of the inverse of a square matrix as $A^{-\top}$.

The reference model is described as
\begin{equation}
    \Dot{x}_m={A_m}x_m+{b_m}r,
    \label{eq2}
\end{equation}
where $x_m$ $\in$ $\mathbb{R}^n$ is the reference state, $r$ $\in$ $\mathbb{R}^q$ is the reference signal, and $A_m$ and $b_m$ are its respectively states matrix and input vector. The dynamics of each agent is represented as
\begin{equation}
    \Dot{x}_i=A_ix_i+(b_i+f_i(x_i))u_i, \hspace{0.5cm} i \in \left[1,...,N\right],
    \label{eq1}
\end{equation}
where $x_i$ $\in$ $\mathbb{R}^n$ are the agent's states, $u_i$ $\in$ $\mathbb{R}^p$ is the control input, $A_i$ is an unknown matrix associated to the agent's states, $b_i$ are known input vectors, and $f_i\colon\,\mathbb{R}^n\to\mathbb{R}^p$ is a bounded input uncertainty that acts as a disturbance. Heterogeneity in the network means that $A_i{\not=}A_j$ and $b_i{\not=}b_j$. 

Considering the characteristics of a conventional MRAC, we define some matching conditions through the following assumptions, based on the general characteristics of control based on the reference model and its extension to distributed systems in acyclic cases \cite{Nguyen2018L,Baldi2019}.

\textbf{Assumption 1.} The vector $k_{ij}^*$ and the scalar $k_{rij}^*$ exist and are defined such that
\begin{align}
    A_i&=A_j+{b_j}k_{ij}^{*\top}, \\
    b_i&={b_j}k_{rij}^* .
    \label{eq4a}
\end{align}
Constants in \eqref{eq4a} are known as coupling matching conditions. 

\textbf{Assumption 2.} The vector $k_{mi}^*$ and the scalar $k_{ri}^*$ exist and satisfy
\begin{align}
    A_m&=A_i+{b_i}k_{mi}^{*\top},\\
    b_m&={b_i}k^*_{ri}.
    \label{eq3}
\end{align}
Constants $k_{mi}^*$ and $k_{ri}^*$ in \eqref{eq3} are known as feedback matching conditions.

\textbf{Assumption 3.} Each agent communicate its input through a communication graph that must be acyclic and contain at least one spanning tree where the leader is connected.

\textbf{Problem:} A network of $N$ agents with dynamics \eqref{eq1}, a model reference \eqref{eq2}, and Assumptions 1-3 verified. The control objective is to achieve that all closed-loop signals must be bounded as $t \xrightarrow{} \infty$ for each agent in the presence of input uncertainty parameters or disconnection of neighboring agents.

\section{Adaptive Control Laws} \label{S3}
In this section, the solutions to the proposed problems is tackled in two directions: the first approach is a robust optimal approximation called adaptive optimal
control modification; the second approach is an
approximation of disturbances by neural networks.

\subsection{Adaptive Optimal Control Modification}
The network composed of heterogeneous agents with disturbances approximated from a robust optimal methodology is presented in this subsection. The following proposition presents the adaptive optimal modification base case \cite{Nguyen2008}.

\textbf{Proposition 1.} Assume that agent $x_1$ with dynamics \eqref{eq1} and the reference \eqref{eq2} are directly connected. It is defined the control law as
\begin{equation}
    u_1=k_{m}x_m+k_{r}r-u_{\text{ad}},
    \label{eq6a}
\end{equation} 
where $k_{m}\in\mathbb{R}^p\times\mathbb{R}^n$ is the constant associated with the reference states, $k_{r}\in\mathbb{R}^p\times\mathbb{R}^q$ is associated with the reference signal and their adaptive laws are given by
\begin{align}
    {\dot{k}^\top_{m}}&=-\text{sgn}\left({k_{ri}}^*\right){\gamma}\: {b^\top_m}P\left(x_1-x_m\right)x_1^\top,\\
    {\dot{k}_{r}}&=-\text{sgn}\left({k_{ri}}^*\right){\gamma}\: {b^\top_m}P\left(x_1-x_m\right)r.
    \label{k1s}
\end{align}

An adaptive gain $\gamma>0$ is defined, and $P$ can be obtained by
\begin{equation}
    PA_m+{A^\top_m}P=-Q, \hspace{0.5cm}Q\succ 0.
    \label{eq8}
\end{equation}

The auxiliary parameter $u_{\text{ad}}\in \mathbb{R}^p$ is an additional control term that allows the suppression of input uncertainty parameters, which satisfies
\begin{equation}
    u_{\text{ad}}={\theta}^\top\phi,
    \label{equad}
\end{equation}
where $\theta \in \mathbb{R}^{n{\times}p}$ is the optimal modification parameter defined as
\begin{equation}
    \dot{\theta}=-{\gamma}\left(\phi{e}^{\top}Pb_1-v\: \phi\: \phi^\top\theta{b}^\top_1PA_m^{-1}b_1\right).
    \label{eq7}
\end{equation}

Also, the map $\phi\colon \mathbb{R}^n \to \mathbb{R}^p$ is a known bounded basis function, $b_1$ is the input vector of the follower agent and $v\!>\!0$ is an adaptive optimal constant. Then, with the controller \eqref{eq6a} synchronization error $e_1=x_1-x_m$ is bounded.

\textit{Proof:} It follows from \cite{Nguyen2018L}.

The disturbance management is defined as $\epsilon$, where its representation is
\begin{equation}
    \epsilon^*(x)=\theta^{*\top}\phi(x)-f(x).
    \label{Iumangement}
\end{equation}
This representation allows uncertainties suppression.

From this approach, the following proposition is made for a distributed perspective.

\textbf{Proposition 2.} Consider a second follower agent with dynamics \eqref{eq1}, which is not directly connected to the reference, with the proposed control law
\begin{equation}
    u_2=k^{\top}_{21}x_1+{k^\top_{m2}}\left(x_2-x_1\right)+k_{r21}u_1-\theta^{\top}_2\phi_2,
    \label{eq10}
\end{equation}
and the adaptive laws
\begin{align}
     {\dot{k}_{21}}^{\top}&=-\text{sgn}\left(k^*_{r2}\right){\gamma}\: {b^{\top}_m}P\left(x_2-x_1\right)x_2^\top,\\
     \label{eq11}
    {\dot{k}_{m2}}^{\top}&=-\text{sgn}\left(k^*_{r2}\right){\gamma}\: {b^{\top}_m}P\left(x_2-x_1\right)\left(x_2-x_1\right)^\top,\\
    {\dot{k}_{r21}}&=-\text{sgn}\left(k^*_{r2}\right){\gamma}\: {b^{\top}_m}P\left(x_2-x_1\right)u_1,\\
    \dot{\theta}_{2}&=-\gamma\left(\phi_2(x_2-x_1)^{\top}Pb_2-v\phi_2\phi^\top_2\theta^\top_2{b^{\top}_2}PA^{-1}_mb_2\right),\\
\end{align}
then the synchronization error $e_{21}=x_2-x_1$ is bounded.

\textit{Proof:} In order to validate the convergence of closed-loop signals in this case, the dynamics of the error are defined starting from 
\begin{equation}
    \dot{e}_{21}=A_me_{21}+b_2\left(\tilde{k}^\top_{21}x_1+\tilde{k}^\top_{m2}e_{21}+\tilde{k}^\top_{r21}u_1-\tilde{\theta}^\top_2\phi_2\right),
    \label{e_d2}
\end{equation}
where $\tilde{k}_{21}=k_{21}-k^*_{21}$; $\tilde{k}_{m2}=k_{m2}-k^*_{m2}$;  $\tilde{k}_{r21}=k_{r21}-k^*_{r21}$; $\tilde{\theta}_2=\theta_2-\theta^*_2$. The following Lyapunov function is considered
\begin{equation}
    V_{21}= e^\top_{21}Pe_{21}+\text{tr}\left(\frac{\tilde{k}^\top_{21}\tilde{k}_{21}}{\gamma|k^*_{r2}|}\right)+\text{tr}\left(\frac{\tilde{k}^\top_{m2}\tilde{k}_{m2}}{\gamma|k^*_{r2}|}\right)+\frac{\tilde{k}^2_{r21}}{\gamma|k^*_{r2}|}+\text{tr}\left(\tilde{\theta}^\top_2\gamma^{-1}\tilde{\theta}_2\right),
    \label{lyap2}
\end{equation}
this equation derived in $\dot{e}_{21}$ is
\begin{align}
    \dot{V}_{21}&=e^\top_{21}\left(PA_m+A^\top_mP\right)e_{21}+2e^\top_{21}Pb_2\left(\tilde{k}^\top_{21}x_1+\tilde{k}^\top_{m2}e_{21}+\tilde{k}^\top_{r21}u_1-\tilde{\theta}^\top_2\phi_2\right)\\
    &+2\text{tr}\left(\frac{\tilde{k}^\top_{21}\gamma^{-1}\dot{\tilde{k}}_{21}}{|k^*_{r2}|}\right)+2\text{tr}\left(\frac{\tilde{k}^\top_{m2}\gamma^{-1}\dot{\tilde{k}}_{m2}}{|k^*_{r2}|}\right)+2\frac{\tilde{k}^\top_{r21}\gamma^{-1}\dot{\tilde{k}}_{r21}}{|k^*_{r2}|}\\
    &-2\text{tr}\left({\tilde\theta}^\top_2\phi_2[e^\top_{21}P-v\phi^\top_2\theta_2b^\top_2PA^{-1}_m]b_2\right),
    \label{dlyap2}
\end{align}
where doing math reduction
\begin{align}
    \dot{V}_{21}&=-e^\top_{21}Qe_{21}+2\left(\text{sgn}\left(k^*_{r2}\right)b^\top_mPe_{21}x^\top_1+\gamma^{-1}\tilde{k}^\top_{21}\right)\frac{\tilde{k}^\top_{21}}{|k^*_{r2}|}\\
    &+2\left(\text{sgn}\left(k^*_{r2}\right)b^\top_mPe_{21}e^\top_{21}+\gamma^{-1}\tilde{k}^\top_{m2}\right)\frac{\tilde{k}^\top_{m2}}{|k^*_{r2}|}\\
    &+2\left(\text{sgn}\left(k^*_{r2}\right)b^\top_mPe_{21}u_1+\gamma^{-1}\tilde{k}^\top_{r21}\right)\frac{\tilde{k}^\top_{r21}}{|k^*_{r2}|}\\
    &+2v\phi^\top_2\tilde{\theta}_2b^\top_2PA^{-1}_mb_2{\tilde{\theta}_2}^\top\phi_2+2v\phi^\top_2\left(\theta^\top_2\phi_2+\epsilon_2\right)b^\top_2PA^{-1}_mb_2{\tilde{\theta}_2}^\top\phi_2.
\end{align}

In this case, we define the condition $b^\top_2PA^{-1}_mb_2<0$ for the equality 
\begin{equation}
    2v\phi^\top_2\tilde{\theta}_2b^\top_2PA^{-1}_mb_2{\tilde{\theta}_2}^\top\phi_2=-v\phi^\top_2\tilde{\theta}_2b^\top_2A^{-\top}_mQA^{-1}_mb_2{\tilde{\theta}_2}^\top\phi_2,
\end{equation}
where $\beta_2\!=\!\lambda_{\min}(b_2A^{-\top}_mQA^{-1}_mb_2)$ and $\beta_3=\frac{\norm{b^{\top}_2PA^{-1}_mb_2}\theta_{02}}{\beta_2}$, $\theta_{02}=\norm{\theta^*_2}$, $\delta_{\epsilon2}=\sup|\epsilon_2|$ is used and can defined the following bound
\begin{equation}
    \norm{e_{21}}\geq\sqrt{\frac{v\beta_2\beta_3\norm{\phi_2}^2}{\beta_1}}=\psi_2,
\end{equation}
therefore, equation \eqref{dlyap2} will be bounded by
\begin{align}
    \dot{V}_{21}&\leq-\lambda_{\min}\left(Q\right)\norm{e_{21}}^2+2\lambda_{\max}(P)\norm{b_2}\left(\norm{\theta^{*\top}_2\phi_2}+\norm{\delta_{\epsilon2}}\right)\norm{e_{21}}\\
    &-{v}\lambda_{min}(Q)\norm{A^{-1}_mb_2\theta^{\top}_2\phi_2}^2,
\end{align}
where $\norm{\theta^{*\top}_2\phi_2}=\norm{\sup_t|\theta^{*\top}_2\phi_2 |}$, in this case, $\phi_2$ is a bounded basis function, then $\norm{e_{21}}$ has as lower bound $\psi_2$, this infers that $\dot{V}_{21}\leq0$, so the background are then met to guarantee that closed loop signals of the agent which are not communicated with the reference are bounded. $\blacksquare$
\\

Continuing with this methodology, it is important to consider different communication links for each agent, since Propositions 1 and 2 only have a link to the leader or another agent, we present the following proposition for an agent with two links.

\textbf{Proposition 3. } Consider a graph $\mathcal{V}=\{1,2,3\}$, $\mathcal{E}=\{(1,2), (2,3)\}$, where the agent $3$ has a link to agents 1 and 2, the control law proposed is
\begin{equation}
    u_3=k^\top_{31}\frac{x_1}{2}+k^\top_{32}\frac{x_2}{2}+k^\top_{m3}\frac{e_{31}+e_{32}}{2}+k^\top_{r31}\frac{u_1}{2}+k^\top_{r32}\frac{u_2}{2}-\frac{\theta^\top_3\phi_3}{2},
    \label{u3aocm}
\end{equation}
where $e_{31}=x_3-x_1$ and $e_{32}=x_3-x_2$. Adaptive laws in this case are defined as
\begin{align}
    {\dot{k}_{31}}^\top&=-\text{sgn}\left(k^*_{r3}\right){\gamma}\: {b^{\top}_m}P\left(e_{31}+e_{32}\right)x_1^\top,\\
    {\dot{k}_{32}}^\top&=-\text{sgn}\left(k^*_{r3}\right){\gamma}\: {b^{\top}_m}P\left(e_{31}+e_{32}\right)x_2^\top,\\
    {\dot{k}_{m3}}^{\top}&=-\text{sgn}\left(k^*_{r3}\right){\gamma}\: {b^{\top}_m}P\left(e_{31}+e_{32}\right)\left(e_{31}+e_{32}\right)^\top,\\
    {\dot{k}_{r31}}&=-\text{sgn}\left(k^*_{r3}\right){\gamma}\: {b^{\top}_m}P\left(e_{31}+e_{32}\right)u_1,\\
    {\dot{k}_{r32}}&=-\text{sgn}\left(k^*_{r3}\right){\gamma}\: {b^{\top}_m}P\left(e_{31}+e_{32}\right)u_2,\\
    \dot{\theta}_{3}&=-\gamma\left(\phi_3(e_{31}+e_{32})^{\top}Pb_3-v\phi_3\phi^\top_3\theta^\top_3{b^{\top}_3}P{A^{-1}_m}b_3\right),
    \label{al3}
\end{align}
then, a synchronization of the errors $e_{31}$ and $e_{32}$ are bounded.

\textit{Proof: } It follows from \cite{Arevalo2019}.

From these formulations, it is possible to define in general a control law that allows the synchronization of an agents network with respect to a reference model and suppressing input uncertainties, the following theorem is then proposed.

\textbf{Theorem 1.} For a $N$ heterogeneous agents network, with dynamics \eqref{eq1}, a reference model \eqref{eq2}, and Assumptions 1-3 hold, defining the following control law
\begin{equation}
    \bar{a}u_i=\bar{a}{k}^\top_{ij}x_i+k_{mi}\Xi_{ij}+\bar{a}k_{rij}u_i-\theta^{\top}_i\phi_i,
    \label{eq12}
\end{equation}
where $\Xi_{ij}=\sum_{j=1}^{N}a_{ij}(x_i-x_j)$ and $\bar{a}=\sum_{j=1}^{N}a_{ij}$. Similarly, with the adaptive laws
\begin{align}
    {\dot{k}^\top_{ij}}=&-\text{sgn}({k^*_{ri}})\gamma\: {b^\top_m} P\Xi_{ij}x^\top_i,\\
    {\dot{k}^\top_{mi}}=&-\text{sgn}({k^*_{ri}})\gamma\: {b^\top_m}P\Xi_{ij}\Xi^\top_{ij},\\
    \dot{k}_{rij}=&-\text{sgn}({k^*_{ri}})\gamma\: {b^\top_m}P\Xi_{ij}u_i.
    \label{alg}
\end{align}
The matched uncertainty parameter is selected from \eqref{equad}, with the optimal modification parameter denoted as
\begin{equation}
    \dot{\theta}_{i}=-\gamma\left(\phi_i(x_i-x_j)^{\top}Pb_i-v\phi_i\phi^\top_i\theta^\top_i{b^\top_i}P{A}^{-1}_mb_i\right).
    \label{eq13a}
\end{equation}
Then, all synchronization error are bounded. 

\textit{Proof: } To validate the boundary of closed-loop signals, the following Lyapunov function is defined
\begin{align}
V&= \sum_{i=1}^{N}\Xi_{ij}^\top P\Xi_{ij}+\sum_{j=1}^{N} \text{tr}\left [ \frac{{\tilde{k}_{mi}}^\top \tilde{k}_{mi}}{\gamma\, \left | {k^*_{ri}} \right | } \right ]+\sum_{i=1}^{N} \bar{a}\; \text{tr}\left [ \frac{{\tilde{k}_{ij}}^\top \tilde{k}_{ij}}{\gamma\, \left | {k^*_{ri}} \right | } \right ]+\sum_{i=1}^{N}\bar{a}\frac{{\tilde{k}_{ri}}^2}{\gamma\, \left | {k^*_r} \right | }\\
&+\sum_{i=1}^{N}\text{tr}(\tilde{\theta}^\top_i\gamma ^{-1}\tilde{\theta}_i),
\label{eqlyap_alt}
\end{align}
where $j=0$ represents the reference model. For the analysis, the dynamics of the error is selected as $e_i=x_i-x_m$ and
\begin{equation}
 \dot{e}_{ij}=A_me_{ij}+b_{i}[{\tilde{k}_{ij}}^{\top}x_i+{\tilde{k}_{mi}}^{\top}e_{ij}+{\tilde{k}_{rij}}^{\top}u_i-\tilde{\theta_i}^\top\phi_i-\tilde{\theta_j}^\top\phi_j].
 \label{e_d}
\end{equation}

With this dynamic, the derivative along its trajectory is
\begin{align}
    \dot{V}&=\sum_{i=1}^N\Xi^\top_{ij}(PA_m+A^\top_mP)\Xi_{ij}+2\Xi^{\top}_{ij}Pb_i\left[\bar{a}{\tilde{k}_{ij}}^{\top}x_i+{\tilde{k}_{mi}}^\top\bar{a}e_{ij}+\bar{a}\tilde{k}_{rij}u_i-\theta^\top_j\phi_j\right]\\
    &+\sum_{i=1}^N{\text{tr}\left[\frac{{\tilde{k}_{mi}}^\top\gamma^{-1}\dot{\tilde{k}}_{mi}}{|k^*_{ri}|}\right]}+\sum_{i=1}^N{\text{tr}\left[\frac{{\tilde{k}_{ij}}^\top\gamma^{-1}\dot{\tilde{k}}_{ij}}{|k^*_{ri}|}\right]}+\sum_{i=1}^N{\bar{a}\frac{\tilde{k}_{rij}\gamma^{-1}\dot{\tilde{k}}_{rij}}{|k^*_{ri}|}}\\
    &-2\sum_{i=1}^N\sum_{j=1}^N\text{tr}\left({\tilde{\theta}_i}^\top\phi_i[e^\top_{ij}P-v\phi^\top_i\theta_ib^\top_iPA^{-1}_m]b_i\right),
    \label{dlyapg}
\end{align}
reducing this Lyapunov function
\begin{align}
    \dot{V}&=-\sum_{i=1}^N\Xi^{\top}_{ij}Q\Xi_{ij}+\sum_{i=1}^N2v\phi^{\top}_i\tilde{\theta}_ib^\top_iPA^{-1}_mb_i{\tilde{\theta}_i}^\top\phi_i\\
    &+\sum_{i=1}^N2v\phi^\top_i\left(\theta^\top_i\phi_i+\epsilon_i\right)b^\top_iPA^{-1}_mb_i{\tilde{\theta}_i}^\top\phi_i.
\end{align}

On the case, $b^\top_iPA^{-1}_mb_i<0$, hence
\begin{equation}
    2v\phi^\top_i\tilde{\theta}_ib^\top_iPA^{-1}_mb_i{\tilde{\theta}_i}^\top\phi_i=-v\phi^\top_i\tilde{\theta}_ib^\top_iA^{-\top}_mQA^{-1}_mb_i{\tilde{\theta}_i}^\top\phi_i.
\end{equation}

Defining the parameters $\beta_1\!=\!\lambda_{\min}(Q)$, $\beta_2\!=\!\lambda_{\min}(b_iA^{-\top}_mQA^{-1}_mb_i)$ and $\beta_3=\frac{\norm{b^{\top}_iPA^{-1}_mb_i}\theta_{0i}}{\beta_2}$, $\theta_{0i}=\norm{\theta^*_i}$ and $\delta_{\epsilon{i}}=\sup|\epsilon_i|$, we define the following bounds
\begin{equation}
    \sum_{i=1}^N\sum_{j=1}^N\norm{e_{ij}}\geq\sqrt{\frac{v\beta_2\beta_3\norm{\phi_i}^2}{\beta_1}}=\psi_i,
\end{equation}
so, \eqref{dlyapg} is bounded by
\begin{align}
    \dot{V}&\leq-\sum_{i=1}^N\lambda_{\min}\left(Q\right)\sum_{j=1}^N\norm{e_{ij}}^2+2\sum_{i=1}^N\sum_{j=1}^N\lambda_{\max}(P)\norm{b_i}\left(\norm{\theta^{*\top}_i\phi_i}+\norm{\delta_{\epsilon{i}}}\right)\norm{e_{ij}}\\
    &-\sum_{i=1}^N{v}\lambda_{min}(Q)\norm{A^{-1}_mb_i\theta^{\top}_i\phi_i}^2,
\end{align}
where $\norm{\theta^{*\top}_i\phi_i}=\norm{\sup_t|\theta^{*\top}_i\phi_i |}$, as $\phi_i$ is a bounded basis function, therefore $\norm{e_{ij}}$ has as lower bound $\psi_i$, this implies that $\dot{V}\leq0$, so we can guarantee that all the synchronization errors $e_{ij}$ are bounded. $\blacksquare$ \\

\textbf{Remark 1.} Compared to similar control strategies for multi-agent systems involving observers \cite{Cai2017}, the proposed scheme only uses the communication of the input between agents, also guaranteeing synchronization under a pre-established reference model.

\subsection{Neural Network Approximation}
In this case, a network of $N$ heterogeneous agents, the input uncertainty, and the Assumptions 1-3 are also used and verified. The following proposition shows the protocol used to synchronize an agent with a reference \cite{Nguyen2008}.

\textbf{Proposition 4.} Consider an agent with dynamics \eqref{eq1} and a reference model \eqref{eq2}, neural network approximation control law for the agent synchronization is
\begin{equation}
u_1={k^\top_{m1}}x_1+k_{r1}r-\theta^{\top}_1\phi_1(W^\top_1\bar{x}_1),
\label{eq6}
\end{equation}
where the uncertainty approximation depends on the function $\phi_1(W^\top_1\bar{x}_1)$. The adaptive laws $k_{
m1}$ and $k_{r1}$ are obtained by \eqref{k1s}, while the adaptive laws associated with neural networks are
\begin{align}
\dot{\theta}_{1}=&-\gamma\phi_1(W^\top_1\bar{x}_1)(x_1-x_0)^{\top}Pb_1,\\
\dot{W}_1=&-\gamma\bar{x}_1(x_1-x_0)^{\top}Pb_1V^\top\sigma(W^\top_1\bar{x}_1),
\label{nn1}
\end{align}
where $\theta_{1}$ and $W_1$ are weight adaptive matrices, $\bar{x}_1=[1\hspace{0.2cm}x^\top_1]^{\top} \in \mathbb{R}^{n+1}$, $V \in \mathbb{R}^{p{\times}n}$ is a bias vector,  $\phi_1(W^\top_1\bar{x}_1)=[1\hspace{0.2cm}\sigma^\top_1(W^\top_1\bar{x}_1)]^\top \in \mathbb{R}^{n+1}$, $a$ is "bell-shaped" approximation constant and with $\sigma_1(x_1)$ as a sigmoidal function described by
\begin{equation}
\sigma_i(x_i)=\frac{1}{1+e^{-ax_i}},
\end{equation}
then, with adaptive protocol \eqref{eq6} synchronization error $e_1$ is bounded.

\textit{Proof: }Its obtained from \cite{Nguyen2008}.

As in the previous case, the distributed extension for agents who do not have communication with the reference is defined in the following proposition.

\textbf{Proposition 5.} Consider a agent with dynamic \eqref{eq1} uncommunicated with the reference \eqref{eq2}, the following control law is proposed
\begin{equation}
    u_2=k^{\top}_{21}x_1+{k^\top_{m2}}\left(x_2-x_1\right)+k_{r21}u_1-\theta^{\top}_2\phi_2(W^\top_2\bar{x}_2),
    \label{eq10nn}
\end{equation}
the adaptive laws associated with the MRAC $k^{\top}_{21},k^\top_{m2},k_{r21}$ are taken from \eqref{eq11}, and the laws associated with neural networks are
\begin{align}
\dot{\theta}_{2}=&-\gamma\phi_2(W^\top_2\bar{x}_2)(x_2-x_1)^{\top}Pb_2,\\
\dot{W}_2=&-\gamma\bar{x}_2(x_2-x_1)^{\top}Pb_2V^\top\sigma(W^\top_2\bar{x}_2),
\label{eq13nn2}
\end{align}
then, the synchronization error $e_{21}$ is bounded.

\textit{Proof: }To ensure boundary of the closed-loop signals, the error dynamics is defined as
\begin{equation}
\dot{e}_{21}=A_me_{21}+b_2[u_2-k^{*\top}_{21}x_1-k^{*\top}_{m2}e_{21}-k^{*\top}_{r21}u_1-\theta^*_2\phi_2+\epsilon^*_2-\theta^*_1\phi_1+\epsilon^*_1],
\label{e_dnn}
\end{equation}
also considering the matching conditions described in Assumptions 1-3, and the Lyapunov function \eqref{lyap2}. The derivative along \eqref{e_dnn} is
\begin{align}
	\dot{V}_{21}&=e_{21}^\top(PA_m+A^\top_mP)e_{21}+2e_{21}^{\top}Pb_2\left[{{\tilde{k}_{21}}^{\top}x_2}+{\tilde{k}_{m2}}^\top{e_{21}}+{\tilde{k}_{r21}u_1}-\theta^\top_2\phi_2+\epsilon^*_2\right]\\
	&+{\text{tr}\left(\frac{{\tilde{k}_{m2}}^\top\gamma^{-1}\dot{\tilde{k}}_{m2}}{|k^*_{r2}|}\right)}+{\text{tr}\left(\frac{{\tilde{k}_{21}}^\top\gamma^{-1}\dot{\tilde{k}}_{21}}{|k^*_{r2}|}\right)}+{{\frac{\tilde{k}_{r21}\gamma^{-1}\dot{\tilde{k}}_{r21}}{|k^*_{r2}|}}}-2\text{tr}\left({\tilde{\theta}_2}^\top\phi_2e^\top_{21}Pb_2\right),
	\label{dlyapg_nn}
	\end{align}
where doing math reduction
\begin{align}
\dot{V}_{21}&=-{e_{21}}^{\top}Q{e_{21}}+2{e_{21}}^{\top}Pb_2\left({\tilde{\theta}_2}^\top\phi_2+\epsilon^*_2\right),
\end{align}
and
\begin{equation}
\dot{V}_{21}=-{e_{21}}^{\top}Q{e_{21}}+2{e_{21}}^{\top}Pb_2\epsilon^*_2\leq-\lambda_{\min}\left(Q\right)\norm{e_{21}}^2+2\norm{Pb_2}\norm{e_{21}}\epsilon^*_0.
\end{equation}

In this case, $\dot{V}\leq0$ if we consider
\begin{equation}
-\lambda_{\min}\left(Q\right)\norm{e_{21}}^2+2\norm{Pb_2}\norm{e_{ij21}}\epsilon^*_0\leq0\Rightarrow\norm{e_{21}}\geq\frac{2\norm{Pb_2}\epsilon^*_0}{\lambda_{\min}Q}.
\end{equation}

With this we can conclude that the synchronization error of an agent that is not directly communicated with the reference using an approximation with neural networks is bounded.
$\blacksquare$

It has been proven that the control law \eqref{eq10nn} allows the synchronization of an agent disconnected from the reference using a protocol approximated by neural networks. From this approach, the control law for an agent with an input connection of two agents is determined, defining a set of coupled error dynamics, as presented in the following proposition.

\textbf{Proposition 6.} Consider an agent with dynamics \eqref{eq1} that is not connected to the reference and has two agents in its vicinity, the following control law is proposed
\begin{equation}
    u_3=k^\top_{31}\frac{x_1}{2}+k^\top_{32}\frac{x_2}{2}+k^\top_{m3}\frac{e_{31}+e_{32}}{2}+k^\top_{r31}\frac{u_1}{2}+k^\top_{r32}\frac{u_2}{2}-\frac{\theta^\top_3\phi_3(W^\top_3\bar{x}_3)}{2},
    \label{u3nn}
\end{equation}
where the adaptive laws $k^\top_{31}, k^\top_{32}, k^\top_{m3}, k^\top_{r31}, k^\top_{r32}$ are taken from \eqref{al3}, and the laws associated with neural networks are
\begin{align}
\dot{\theta}_{3}=&-\gamma\phi_3(W^\top_3\bar{x}_3)(e_+{31}+e_{32})^{\top}Pb_3,\\
\dot{W}_3=&-\gamma\bar{x}_3(e_+{31}+e_{32})^{\top}Pb_3V^\top\sigma(W^\top_3\bar{x}_3).
\label{eq13nn}
\end{align}
Then, with this control law, synchronization errors $e_{31}$ and $e_{32}$ are bounded.

\textit{Proof: }In this case, the proof takes the same structure as the Proposition 5.

Finally, with this methodological description, the following theorem presents in a general way the control law and its capability for handling input uncertainties in heterogeneous systems through neural networks.

\textbf{Theorem 2.} Considering a $N$ heterogeneous network with dynamics \eqref{eq1} and a reference \eqref{eq2}, for agents that are not directly communicated with the reference, the control law with neural network approximation used for synchronization is
\begin{equation}
\bar{a}u_i=\bar{a}{{k}^\top_{mij}}x_j+k_{mi}\bar{\Xi}_{ij}+k_{rij}\bar{a}u_j-\theta^{\top}_i\phi_i(W^\top_i\bar{x}_i).
\label{u_nn}
\end{equation}

The adaptive laws associated with MRAC are \eqref{alg}, and the adaptive control laws associated with neural networks are
\begin{align}
\dot{\theta}_{i}=&-\gamma\phi_i(W^\top_i\bar{x}_i)e_{ij}^{\top}Pb_i,\\
\dot{W}_i=&-\gamma\bar{x}_ie_{ij}^{\top}Pb_iV^\top\sigma(W^\top_i\bar{x}_i),
\label{eq13}
\end{align}
then the control law \eqref{u_nn} allow the agents to synchronize its dynamics with respect to the reference model.

\textit{Proof: } This proof validates the general bounding of closed-loop synchronization errors in the presence of input uncertainties for heterogeneous agents, the dynamics of the error in this case is defined as
\begin{equation}
\dot{e}_{ij}=A_me_{ij}+b_i[u_i-k^{*\top}_{mij}x_j-k^{*\top}_{mi}e_{ij}-k^{*\top}_{rij}u_j-\theta^*_i\phi_i-\epsilon^*_i+\theta^*_j\phi_j+\epsilon^*_j],
\label{e_dnnt}
\end{equation}
where considering $\tilde{k}_{mi}=k_{mi}-k^*_{mi}$; $\tilde{k}_{mij}=k_{mij}-k^*_{mij}$;
$\tilde{k}_{ri}=k_{ri}-k^*_{ri}$; $\tilde{k}_{rij}=k_{rij}-k^*_{rij}$; $\tilde{\theta}_i=\theta_i-\theta^*_i$, the Lyapunov function considering in this case is
\begin{align}
V&=\sum_{i=1}^{N}\Xi_{ij}^\top P\Xi_{ij}+\sum_{j=1}^{N} \text{tr}\left ( \frac{{\tilde{k}_{mi}}^\top \tilde{k}_{mi}}{\gamma\, \left | {k^*_{ri}} \right | } \right )+\sum_{i=1}^{N} \bar{a}\; \text{tr}\left ( \frac{{\tilde{k}_{mij}}^\top \tilde{k}_{mij}}{\gamma\, \left | {k^*_{ri}} \right | } \right )+\sum_{i=1}^{N}\bar{a}\frac{{\tilde{k}_{ri}}^2}{\gamma\, \left | {k^*_r} \right | }\\
&+\text{tr}(\tilde{\theta}^\top_i\gamma ^{-1}\tilde{\theta}_i).
\label{eqlyap}
\end{align}

The derivative of \eqref{eqlyap} along \eqref{e_dnnt} is
\begin{align}
	\dot{V}&=\sum_{i=1}^N\Xi_{ij}^\top(PA_0+A^\top_0P)\Xi_{ij}\\
	&+2\Xi_{ij}^{\top}Pb_i\left[\bar{a}{\tilde{k}_{mij}}^{\top}x_i+{\tilde{k}_{mi}}^\top\Xi_{ij}+\bar{a}\tilde{k}_{rij}u_i-\theta^\top_i\phi_i+\epsilon^*_i\right]\\
	&+\sum_{i=1}^N{\text{tr}\left(\frac{{\tilde{k}_{mi}}^\top\gamma^{-1}\dot{\tilde{k}}_{mi}}{|k^*_{ri}|}\right)}+\sum_{i=1}^N{\text{tr}\left(\frac{{\tilde{k}_{mij}}^\top\gamma^{-1}\dot{\tilde{k}}_{ij}}{|k^*_{ri}|}\right)}+\sum_{i=1}^N{\bar{a}{\frac{\tilde{k}_{rij}\gamma^{-1}\dot{\tilde{k}}_{rij}}{|k^*_{ri}|}}}\\
	&-2\sum_{i=1}^N\sum_{j=1}^N\text{tr}\left({\tilde{\theta}_i}^\top\phi_ie^\top_{ij}Pb_i\right),
	\label{dlyapg2}
	\end{align}
also doing mathematical reduction, we find that
\begin{align}
\dot{V}&=-\sum_{i=1}^N\Xi_{ij}^{\top}Q\Xi_{ij}+2\Xi_{ij}^{\top}Pb_i\left({\tilde{\theta}_i}^\top\phi_i+\epsilon^*_i\right),
\end{align}
then
\begin{equation}
\dot{V}=-\sum_{i=1}^N\Xi_{ij}^{\top}Q\Xi_{ij}+2\Xi_{ij}^{\top}Pb_i\epsilon^*_i\leq-\sum_{i=1}^N\lambda_{\min}\left(Q\right)\sum_{j=1}^N\norm{e_{ij}}^2+2\sum_{i=1}^N\sum_{j=1}^N\norm{Pb_i}\norm{e_{ij}}\epsilon^*_0.
\end{equation}

Validating, $\dot{V}\leq0$ if
\begin{equation}
-\sum_{i=1}^N\lambda_{\min}\left(Q\right)\sum_{j=1}^N\norm{e_{ij}}^2+2\sum_{i=1}^N\sum_{j=1}^N\norm{Pb_i}\norm{e_{ij}}\epsilon^*_0\leq0\Rightarrow\Rightarrow\sum_{i=1}^N\sum_{j=1}^N\norm{e_{ij}}\geq\frac{2\norm{Pb_i}\epsilon^*_0}{\lambda_{\min}Q},
\end{equation}
this inequality makes it possible to ensure that closed-loop synchronization errors are bounded in a heterogeneous network with input uncertainty worked from a neural network approach. $\blacksquare$

With these control laws, it is possible to validate a robust approach for a multi-agent control in the presence of input uncertainty parameters. Next, we show a second case in this type of network associated to the input estimation in disconnected agents.

\section{Input Estimation} \label{S4}
In this section, we present the case where some neighboring agent $j$ cannot communicate the input value to its neighborhood. The definition of a control law that allows the estimation of neighbors inputs in order to avoid failures in the proposed control laws due to the communication is presented. The control law to employ is
\begin{equation}
\bar{a}u_i=\bar{a}{{k}^\top_{ij}}x_i+k_{mi}\Xi_{ij}+\bar{a}\hat{u}_{ji}-\theta^{\top}_i\phi_i,
\label{u_ie}
\end{equation}
The adaptive laws associated with the MRAC are taken from \eqref{alg}, validating that the law $k_{rij}$ is no longer used due to the isolation of the input from neighboring agents, the control law for the input estimation $\hat{u}_{ji}$ is
\begin{equation}
\dot{\hat{u}}_{ji}=-\text{sgn}(k_{ri}^*)\gamma{b_0^\top}P\Xi_{ij}.
\label{eq16}
\end{equation}

\textbf{Theorem 3. } Considering a network of $N$ heterogeneous agents with dynamics \eqref{eq1} and a reference model \eqref{eq2}. Then, controller \eqref{u_ie} allow the synchronization of the network even in the presence of agents isolated to its neighborhood.

\textit{Proof: }This proof also validates the closed-loop synchronization error boundary. In this case, the dynamic of the error is defined as
\begin{equation}
\dot{e}_{ij}=A_me_{ij}+b_i[u_i-k^{*\top}_{mij}x_j-k^{*\top}_{mi}e_{ij}-u^*_{ji}-\theta^*_i\phi_i-\epsilon^*_i+\theta^*_j\phi_j+\epsilon^*_j],
\label{e_diu}
\end{equation}
where  $\tilde{u}_{ji}=u_{ji}-u^*_{ji}$, we can define the following Lyapunov equation
\begin{align}
V&=\sum_{i=1}^{N}\Xi_{ij}^\top P\Xi_{ij}+\sum_{j=1}^{N} \text{tr}\left ( \frac{{\tilde{k}_{mi}}^\top \tilde{k}_{mi}}{\gamma\, \left | {k^*_{ri}} \right | } \right )+\sum_{i=1}^{N} \bar{a}\; \text{tr}\left ( \frac{{\tilde{k}_{mij}}^\top \tilde{k}_{mij}}{\gamma\, \left | {k^*_{ri}} \right | } \right )+\sum_{i=1}^{N}\bar{a}\frac{{\tilde{u}_{ji}}^2}{\gamma\, \left | {k^*_r} \right | }\\
&+\text{tr}(\tilde{\theta}^\top_i\gamma ^{-1}\tilde{\theta}_i),
\label{eqlyap_iu}
\end{align}
where its derivative along \eqref{e_diu} is
\begin{align}
	\dot{V}&=\sum_{i=1}^N\Xi_{ij}^\top(PA_0+A^\top_0P)\Xi_{ij}\\
	&+2\left[\Xi_{ij}\right]^{\top}Pb_i\left[\bar{a}\tilde{k}_{mij}^{\top}x_i+{\tilde{k}_{mi}}^\top\Xi_{ij}+\bar{a}\tilde{u}_{ji}-\theta^\top_i\phi_i+\epsilon^*_i\right]+\sum_{i=1}^N{\text{tr}\left(\frac{{\tilde{k}_{mi}}^\top\gamma^{-1}\dot{\tilde{k}}_{mi}}{|k^*_{ri}|}\right)}\\
	&+\sum_{i=1}^N{\text{tr}\left(\frac{{\tilde{k}_{mij}}^\top\gamma^{-1}\dot{\tilde{k}}_{ij}}{|k^*_{ri}|}\right)}+\sum_{i=1}^N{\bar{a}\frac{\tilde{u}_{ji}\gamma^{-1}\dot{\tilde{u}}_{ji}}{|k^*_{ri}|}}-2\sum_{i=1}^N\sum_{j=1}^N\text{tr}\left({\tilde{\theta}_i}^\top\phi_ie^\top_{ij}Pb_i\right),
	\label{dlyapg_a}
	\end{align}
and doing math reduction,
\begin{align}
\dot{V}&=-\sum_{i=1}^N\Xi_{ij}^{\top}Q\Xi_{ij}+2\Xi_{ij}^{\top}Pb_i\left({\tilde{\theta}_i}^\top\phi_i+\epsilon^*_i\right),
\end{align}
then
\begin{equation}
\dot{V}=-\sum_{i=1}^N\Xi_{ij}^{\top}Q\Xi_{ij}+2\Xi_{ij}^{\top}Pb_i\epsilon^*_i\leq-\sum_{i=1}^N\lambda_{\min}\left(Q\right)\sum_{j=1}^N\norm{e_{ij}}^2+2\sum_{i=1}^N\sum_{j=1}^N\norm{Pb_i}\norm{e_{ij}}\epsilon^*_0.
\end{equation}

In this case, in the same way $\dot{V}\leq0$ if
\begin{align}
-&\sum_{i=1}^N\lambda_{\min}\left(Q\right)\sum_{j=1}^N\norm{e_{ij}}^2+2\sum_{i=1}^N\sum_{j=1}^N\norm{Pb_i}\norm{e_{ij}}\epsilon^*_0\leq0\Rightarrow\Rightarrow\sum_{i=1}^N\sum_{j=1}^N\norm{e_{ij}}\geq\frac{2\norm{Pb_i}\epsilon^*_0}{\lambda_{\min}Q},
\end{align}
with this condition we can then ensure that all synchronization errors are bounded even when there is no communication of the input between agents. $\blacksquare$

\section{Simulation Results} \label{S5}
This section presents the study case of cooperative cruise control as an application of proposed Theorems 1-3 and the simulation results obtained.

\subsection{Motivational Example}
\begin{figure}[ht]
	\centering
	\includegraphics[width=0.48 \textwidth]{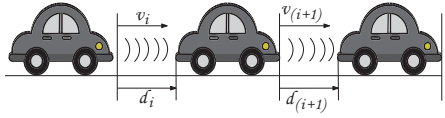}
	\caption{Cooperative cruise control methodology.}	
	\label{fig:carros}
\end{figure}

We introduce the cooperative cruise control, where a network of vehicles will maintain the same speed and distance between each other. The technique is known as cooperative adaptive cruise control. Figure \ref{fig:carros} shows the graphic interpretation of this methodology, where $v_i$ represents the speed present in each vehicle and $d_i$ the distance between them. Each agent is modeled as
\begin{eqnarray}
 \dot{x}_i = \left[ \begin{array}{ccc}
0 & 1 & 0 \\
0 & 0 & 1 \\
0 & 0 & -\frac{1}{\tau_i} \end{array} \right]\
x_i+ \left(\left[ \begin{array}{c}
0 \\
0 \\
\frac{1}{\tau_i} \end{array} \right] + f_i(x_i)\right) u_i,
\label{eq17}
\end{eqnarray}
where $\tau_i$ represents the inertial time lag of the power-train system, that varies between each vehicle, according to its physical characteristics and its environment. Input $u_i$ is the acceleration defined as the force per vehicle mass. In this case, there is a leading agent that defines the acceleration profile that the vehicle network must maintain. If the vehicles only maintain internal sensors in each one, it would be just a case of adaptive cruise control, where each vehicle in their neighborhood must take a measurement of the speed and position of them, to take the pertinent actions, under the logic of string stability \cite{Besselink2017}. In cooperative adaptive cruise control, vehicles come with a communication technology via wireless sensors incorporated, allowing better operation at the network level. The type of control laws shown in this work have a great contribution derived from the fact that most of the related works to date have focused on the control of homogeneous systems, without considering that on highways, there is prominence of heterogeneity in the network \cite{Yang2016}.

\subsection{Simulations}
\begin{figure}[ht]
	\centering
	\includegraphics[width = 0.2\textwidth]{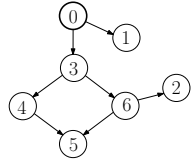}
	\caption{Leader-follower communication graph.}	
	\label{fig:1}
\end{figure}

\begin{figure}[htbp]
	\centering
	\includegraphics[width = 0.5\textwidth]{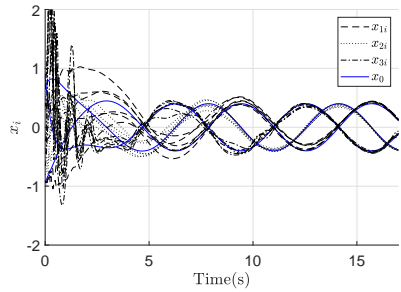}
	\caption{Temporal response agents synchronization with optimal adaptive law.}	
	\label{fig:aocm}
\end{figure}

\begin{figure}[htbp]
	\centering
	\includegraphics[width = 0.5\textwidth]{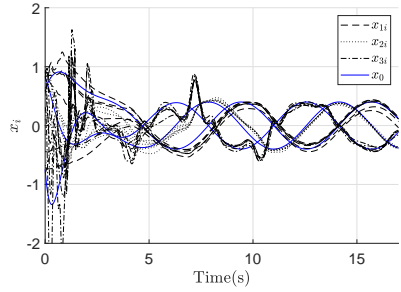}
	\caption{Temporal response agents synchronization with neural network approximation.}	
	\label{fig:nn}
\end{figure}

Considering a network of vehicles, we define the communication graph in Figure \ref{fig:1} to validate the algorithms developed.
In this case, the dynamics of each agent take the form of \eqref{eq17}, where the coefficients are observed in Table 1.
\begin{table}[htbp]
	\vspace{0.3cm}
	\centering
	\caption{Agent's Coefficients and Initial Conditions}
	\begin{tabular}{ccc}
		\hline
		& $\tau$ & $x_0$ \\ 
		\hline
		$A_0$ & -4 & $[1 \hspace{0.1cm} -1]^\top$ \\  
		$A_1$ & 1 & $[1 \hspace{0.1cm} 0]^\top$ \\ 
		$A_2$ & 0.4 & $[-1 \hspace{0.1cm} 0.5]^\top$ \\ 
		$A_3$ & 0.25 & $[1 \hspace{0.1cm}  0]^\top$ \\ 
		$A_4$ & 0.45 & $[-1 \hspace{0.1cm}  1]^\top$ \\ 
		$A_5$ & 0.5 & $[-0.5 \hspace{0.1cm}  1]^\top$ \\ 
		$A_6$ & 1.25 & $[0 \hspace{0.1cm}  -1]^\top$ \\ \hline
	\end{tabular}
	\label{tab:my_label}
\end{table}
The agent $0$ acts as the reference model, which is considered as the only stable open-loop system. For the development of adaptive laws, the following parameters are necessary $\gamma = 10$, $v=1$ and $Q=\text{diag}(10,1,1)$ is the matrix associated with the solution of the linear Lyapunov function. The matching conditions associated with the neighbors and the reference are initialized to 0. The input uncertainty defined for simulation is $f_i(x_i)=0.1\sin{x_{3i}}$. Figure \ref{fig:aocm} shows the response of the system with adaptive optimal control law included, with a reference value $r=2\sin(t) $ and it is observed that in the presence of uncertainties the desired behavior is achieved. Similarly, Figure \ref{fig:nn} shows the temporal response of the same agents network with input uncertainty but managed through the neural network approximation, we can observe that this approximation also allows to suppress the included disturbance with a slight oscillation in its temporal response.

\begin{figure}[htbp]
	\centering
	\includegraphics[width = 0.5\textwidth]{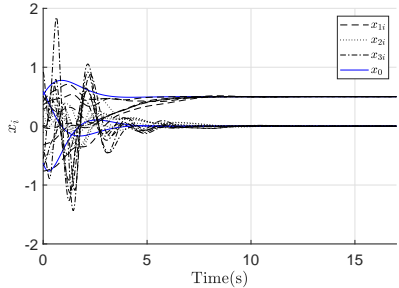}
	\caption{Temporal response with input estimation for uncommunicated agents synchronization.}	
	\label{fig:ie}
\end{figure}

Finally, the result in the network is validated through the input estimation for agents who cannot communicate it, the response in this case is developed and its response is observed in Figure \ref{fig:ie} with a constant reference $r=2$ instead of a sinusoidal reference due to fluctuations that may present.

\section{Conclusions} \label{S6}
In this work, adaptive control laws based on robust models are presented in the case of cooperative cruise control study. An adaptive optimal control law and a neural network based approach are proposed for the suppression of each agent input uncertainty parameters. Likewise, an estimation law is used for its application in the case of disconnected agents. For the adaptive control laws the use of matching conditions allows a synchronization of the agents with the reference and its neighbors. In the case of uncertainty, these conditions must be estimated. 
The theoretical results obtained make it possible to guarantee the limitation of synchronization errors in the closed-loop network under pre-established conditions.
The cooperative adaptive cruise control application case, being a platoon synchronization methodology through input communication and being extensively studied in the literature, can be shown as the motivational example of these control laws.
Future developments may focus on extending algorithms to nonlinear systems or with switched topologies in an output regulation methodology.

\bibliography{main}

\end{document}